# Quantum Size Effects on the Chemical Sensing Performance of Two-Dimensional Semiconductors

Junghyo Nah[1,2,3,†], S. Bala Kumar[4], Hui Fang[1,2,3], Yu-Ze Chen[5], Elena Plis[6], Yu-Lun Chueh[5], Sanjay Krishna[6], Jing Guo[4] and Ali Javey[1,2,3,*]

[1] *Electrical Engineering and Computer Sciences, University of California, Berkeley, CA, 94720*

[2] *Materials Sciences Division, Lawrence Berkeley National Laboratory, Berkeley, CA 94720*

[3] *Berkeley Sensor and Actuator Center, University of California, Berkeley, CA, 94720*

[4] *Electrical and Computer Engineering, University of Florida, Gainesville, FL, 32611*

[5] *Materials Science and Engineering, National Tsing Hua University, Hsinchu 30013, Taiwan, ROC, 30013*

[6] *Electrical and Computer Engineering, University of New Mexico, Albuquerque, NM, 87106*

\* *Corresponding author: ajavey@eecs.berkeley.edu*

[†] *Present Address: Electrical Engineering, Chungnam National University, Daejeon, 305-764, South Korea*

**Abstract**

We investigate the role of quantum confinement on the performance of gas sensors based on two-dimensional InAs membranes. Pd-decorated InAs membranes configured as $H_2$ sensors are shown to exhibit strong thickness dependence, with ~100x enhancement in the sensor response as the thickness is reduced from 48 to 8 nm. Through detailed experiments and modeling, the thickness scaling trend is attributed to the quantization of electrons which favorably alters both the position and the transport properties of charge carriers; thus making them more susceptible to surface phenomena.

## 1. Introduction

Nanoscale size effects drastically tailor the fundamental physical and chemical behavior of materials; thereby, enabling a wide range of new devices with highly intriguing properties. For instance, nanomaterials have been widely investigated for high performance chemical and biological sensors with higher sensitivity and lower detection limits as compared to their bulk material counterparts.[1,2,3,4,5,6,7,8] Several studies have been made to better understand the various size effects in sensor response.[9,10,11,12,13,14,15,16] In general, enhanced sensor response in nanomaterial-based gas sensors is attributed to the high surface-to-volume ratio. It was also reported that sensor response is sharply increased as the size of the materials approaches its Debye length. Here, through detailed experiments and simulations, we explore for the first time the drastic role of carrier *quantum confinement* on the performance of chemical sensors based on two-dimensional (2-D) semiconductors. The results serve as a general guideline for exploring a novel class of highly sensitive sensors that utilize structural quantization of carriers for improved performance.

## 2. Experimental methods

In this study, InAs membranes with nanoscale-thicknesses (8-48 nm) are used as a model material system. Because of the large Bohr radius of bulk InAs (~34 nm), heavy quantum confinement of electrons is readily observed for sub-20 nm thick membranes.[17] This is in distinct contrast to Si-based devices where the Bohr radius is ~3 nm, and structural confinement can only be observed for an order of magnitude smaller dimensions as compared to InAs. For instance, our recent experimental studies have shown that ~8 nm thick InAs quantum membranes (QMs) exhibit 2-D subbands spacing of >300 meV, with the electrons populating only one subband at room temperature for relatively moderate carrier concentrations.[17] Thereby, sensor response can be systematically studied as a function of InAs thickness in order to shed light on the quantum confinement effects.

InAs QMs were configured into $H_2$ sensors by functionalizing the top surface of the QMs with Pd nanoparticles (NPs). Upon exposure to molecular hydrogen, the work function of Pd is reduced[18,19,20], resulting in the perturbation of electron transport in the InAs QMs. The sensors are reversible with the devices fully recovering their initial electrical characteristics upon the removal of $H_2$. Our results show that InAs thickness scaling on sensor performance serves two roles. First, electrostatic control of carrier concentrations by Pd-functionalized surface reaction with $H_2$ is enhanced with thickness scaling, mainly due to the reduced density of states (DOS) for 2-D semiconductors. Second and more importantly, quantization effects significantly alter the electron transport properties of the devices with the charge centroid moving closer to the top surface of the device. This results in enhanced surface scattering rates of electrons in quantum confined membranes, thus making them more susceptible to surface phenomena.

## 3. Results and Discussion

Figure 1(a) demonstrates the cross-sectional schematic of InAs QM gas sensors. Different thicknesses of InAs QMs (single crystalline) were transferred onto $p^+Si/SiO_2$ (50nm thick, thermally grown) substrates using an epitaxial layer transfer technique previously reported.[21] Figure 1(b) shows atomic force microscope (AFM) images of the transferred InAs QMs having three different thicknesses of $T_{InAs}$ = 8, 18, and 48 nm (from top to bottom), where the native oxide thickness is ~ 2 nm, confirmed by transmission electron microscopy (TEM).[21] The source (*S*) and drain (*D*) contacts were photolithographically defined, followed by Ti (5 nm)/ Au (40 nm) electron-beam evaporation. The heavily doped Si substrate is used as the global back-gate with a 50 nm thick $SiO_2$ as the gate dielectric. The fabricated devices have a channel length of ~2.5 μm and width of ~360 nm. The Pd evaporation (~ 2 nm thickness) on the exposed InAs surface finalized the sensor fabrication. The electrical characteristics change after Pd functionalization is shown in Fig. S1. Figure 1(c) and (d) shows the TEM micrograph of a Pd-coated InAs QM (thickness, 8 nm). The evaporated Pd is aggregated into NPs with an average diameter of ~ 10 nm (Fig. S2).

Figures 2(a)-(c) show the $I_d$-$V_{gs}$ characteristics of the Pd-coated QM devices with $T_{InAs}$ = 8, 18, 48 nm, measured before and after exposure to 0.5 % $H_2$ (diluted in dry air). We note that the devices were exposed to $H_2$ for 10 min, which is sufficient time to reach the saturation levels of the gas sensor responses (*i.e.*, equilibrium state). The conductance change upon $H_2$ exposure is due to a combination of (i) threshold voltage change ($\Delta V_t$) and (ii) field-effect mobility change *($\Delta \mu_{FE}$)* as evident from the change in the slope (i.e., transconductance). Both effects are caused by the lowering of the Pd nanoparticle work function upon $H_2$ exposure, as illustrated in Fig. 3(a). Namely, the carrier concentration (*n*) in InAs QMs is modulated upon the change of Pd nanoparticle work function, resulting in the observed $\Delta V_t$. In addition, randomly deposited Pd particles and their potential fluctuations in $H_2$ cause uneven electric potentials which influence the electron

transport in InAs QMs by altering the scattering rates and thereby mobility. By lowering the work function of Pd NPs, the barrier height for the scattering events is reduced, resulting in increased $\mu_{FE}$.

Clearly, the electrical characteristics of thinner QM devices are more dramatically affected by $H_2$ exposure (Fig. 2). Figure 3(b) shows the sensor responses plotted as a function of $T_{InAs}$, calculated at $V_{gs}$ that the highest sensor response could be obtained in each device. Here, the sensor response is defined as $\Delta G/G_0$, where $G_0$ is the initial conductance in ambient air and $\Delta G$ is the change of the conductance after $H_2$ exposure. The observed $\Delta G/G_0$ is ~ 100× higher for $T_{InAs}$ = 8 nm device as compared to the 48 nm-thick device (Fig. 3b). We note that the sensors are fully reversible, showing consistent response through multiple cycles of gas exposure (Fig. S3 and S4). Figure 3(c) summarizes $\Delta\mu_{FE}$ and $\Delta V_t$ for different InAs QM thicknesses. Two observations are apparent from the data. First, $\Delta V_t$ gradually increases with thickness miniaturization. Specifically, $\Delta V_t$ of ~1.75 and 2.5V are observed for $T_{InAs}$=48 and 8 nm devices, respectively. This result indicates that an enhanced electrostatic control of the InAs channel by Pd particles is achieved in the thinner QMs. Second, the electron mobility change upon $H_2$ exposure exhibits strong thickness dependence with $\Delta\mu_{FE}$ of ~ 400 and 0 $cm^2V^{-1}s^{-1}$ for $T_{InAs}$= 8 and 48 nm devices, respectively. This observation reveals that electron transport in ultrathin QMs is strongly affected by surface effects, which is consistent with the previous reports of the enhanced surface scattering rates in thinner QMs.

Numerical simulation methods were employed to further investigate the effects of InAs QM thickness on the sensor response. Electrostatic potential and carrier concentration across the different InAs thicknesses were obtained by using a self-consistent Poisson-Schrödinger solver.[22] The effect of Pd and $H_2$ are modeled by changing the boundary conditions as explained in the Supplementary Information. Figure 4 summarizes the simulation results of two different InAs thicknesses, 8 nm and 48 nm. Figures 4 (a) and (b) show the calculated energy band diagrams of each thickness before/after $H_2$ exposure. It should be noted that only one subband is being populated in the 8 nm-thick InAs, while the number of populated subbands in the 48 nm-thick InAs depends on the $H_2$ content (i.e., surface potential). Figure 4(c) and (d) demonstrate the carrier concentration profiles along the depth of each InAs QM before (black solid lines) and after (red solid lines) $H_2$ exposure. Due to strong structural quantization in the 8nm-thick InAs, the charge centroid is located close to the center of the channel thickness. On the other hand, in the 48nm-thich InAs, the charge centroid is located close to the InAs/SiO$_2$ bottom interface (i.e., close to the back-gate). Specifically, the charge centroids are ~5 and ~44 nm away from the top surface for $T_{InAs}$ = 8 and 48 nm, respectively. Thus, the carrier concentration modulation due to the change of top surface potential is more effective for the 8nm-thick InAs. As a result, larger $\Delta V_t$ upon $H_2$ exposure is achieved in the 8 nm-thick InAs devices as compared to thicker QMs, consistent with the experimental results (Fig. 4(e)-(f), Fig. S5).

The random distribution of the Pd particles on the InAs surface also perturbs the potential inside the InAs body and hence increases the electron scattering, especially near the top surface of the InAs. To investigate this scattering event, we employed a model similar to the surface roughness scattering model[23]. In our model, the perturbation potential change is defined as $\Delta \Phi_s = |\Phi_2 - \Phi_1|$, where $\Phi_2$ and $\Phi_1$ are the potential across the depth of the InAs with and without $H_2$ exposure, respectively. Thus, $\Delta\Phi_s$ behaves as a sensing potential, which quantifies how the potential perturbation on the top InAs surface caused by $H_2$ exposure extends across the InAs body. Based on extracted $\Delta\Phi_s$ values, we compute the potential-disorder-scattering (PDS) mobility, $\mu_{PDS}$ (see Supplementary Information for the details). The PDS mobility is then employed to gauge the scattering induced electrical characteristics changes. Thus, the total mobility ($\mu$) is simplified as $\mu \approx \mu_{PDS}$. Using this value, the $I_d$ can be qualitatively estimated as $I_d \sim n\mu$, where $n$ is the total charge density in the channel, obtained by using a self-consistent Poisson-Schrödinger solver.

The right axes in Figures 4(c) and (d) show $\Delta\Phi_s$ across the InAs bodies after $H_2$ exposure. The $\Delta\Phi_s$ attenuates much slower in the 8 nm-thick InAs compared with the 48 nm-thick InAs. As a result, in the 8 nm-thick InAs, potential perturbation on the top surface has a more significant effect across the entire body of the channel. This effect can be explained as follows: the 8 nm-thick InAs QM is highly quantized; and due to the effect of quantum confinement, DOS is reduced. Therefore, the screening length is further extended such that potential disorder on the top InAs surface strongly affects electron transport across the entire body. This makes the quantum confined membrane, i.e. the 8 nm-thick InAs QM, more susceptible to surface perturbations. Besides the $\Delta\Phi_s$ attenuation, the overlap between the magnitude of $\Delta\Phi_s$ and the electron concentration is much higher in the 8 nm-thick InAs compared with the 48 nm-thick InAs. This implies that the perturbation potential due to the surface Pd particles has a larger effect on the average carrier in the channel for the 8 nm-thick InAs, and hence achieving a higher sensor response.

The effect of electron scattering due to the surface perturbation potential is reflected in the simulated $I_d$-$V_{gs}$ characteristics as shown in Figures 4(e) and (f). Here we compare the effect before and after $H_2$ exposure for each InAs thickness. Due to the significant (insignificant) change in mobility, the slope of $I_d$-$V_{gs}$ curve for the 8 nm-thick InAs (48 nm-thick InAs) device is increased (remain almost constant) when the device is exposed to $H_2$. This leads to a higher sensor response in the 8 nm-thick InAs. These results qualitatively explain the change in the slope and threshold voltage of the experimental results as shown in Fig. 2.

To highlight the importance of the quantum confinement effects in the sensor response, the above quantum simulation results were compared with those obtained from simulation of the 8nm-thick-InAs QM by assuming hypothetical semiclassical carrier statistics (Supp. Info., Fig. S6). Our results show that the perturbation potential is higher in the quantum simulation as compared to the semiclassical simulation, due to the lower DOS. In addition, the charge centroid is located closer to the top surface in the quantum simulation, thereby resulting in a higher sensitivity of carrier scattering rates to the surface effects. Thus, quantum confinement of electrons amplifies the size effects of the sensor response. This conclusion is further verified by comparing the quantum simulation results of InAs QMs with those of Si QMs of similar thickness (Supp. Info., Fig. S7). We find that the $\Delta\Phi_s$ is more prominent in InAs because of its low effective mass of $0.03m_0$, which leads to larger quantization effects as compared to Si.

It should be noted that besides the effects discussed above, the probability of electron wavefunction leaking out of the surface of InAs is also increased with thickness miniaturization as depicted in Figures 4c-d. This yet presents another mechanism that contributes to the observed enhancement of sensor response as the InAs XOI thickness is reduced. Specifically, tor the 8nm film, the simulation results indicate that electron wavefunction probability of leaking out of the surface is ~0.3%. For comparison, the same probability is ~0.1% for the 48 nm thick XOI film. However, this effect alone does not explain the orders of magnitude improvement of sensitivity obtained.

## 4. Conclusions

In summary, we have experimentally and theoretically demonstrated the drastic role of quantum size effects in the response of chemical sensors, using Pd-decorated InAs QM devices as a model device. We have observed up to two orders higher sensor responses in an 8 nm-thick InAs QM gas sensor by comparison to a 48 nm-thick InAs device. The InAs thickness scaling down to quantum confinement limits enables the enhanced potential modulation as well as significant mobility change upon $H_2$ exposure, a contributing factor to obtaining high sensor response. Notably, while structural quantum confinement of carriers is shown in the past to degrade the transistor performances due to enhanced surface scattering rates[17], here, the same effect is shown to be highly beneficial in enhancing the performance of chemical sensors. The work here could presents the fundamental device physics of chemical sensors based on 2-D semiconductors.

**Acknowledgements**

The device aspects of this work were funded by FCRP/MSD. The materials characterization part of this work was supported by the Director, Office of Science, Office of Basic Energy Sciences, Materials Sciences and Engineering Division, of the U.S. Department of Energy under Contract No. DE-AC02-05CH11231. A.J. acknowledges a Sloan Research Fellowship, NSF CAREER Award, and support from the World Class University program at Sunchon National University. Y.-L.C. acknowledges support from the National Science Council, Taiwan, through grant no. NSC 98-2112-M-007-025-MY3.

**Supporting Information**

Transfer characteristics change of InAs XOI FETs after Pd functionalization; TEM of Pd nanoparticles; time response sensing measurements; details of the electrostatics simulations used for InAs XOI gas sensors. This material is available free of charge via the Internet at http://pubs.acs.org.

**Figure Captions**

**Figure 1. Palladium-decorated InAs QM hydrogen gas sensors.** (a) Schematic representation of an InAs QM device used for the sensing studies. (b) AFM images of transferred InAs QMs on Si/SiO$_2$ substrates prior to Pd deposition, with $T_{InAs}$ = 8, 18, 48 nm from top to bottom. (c) and (d) Cross-sectional TEM images of a 8nm-thick InAs QM decorated with Pd nanoparticles.

**Figure 2. H$_2$ sensing response of InAs QM devices**. Transfer characteristics of Pd-decorated QM devices with $T_{InAs}$ of (a) 8 nm, (b) 18 nm and (c) 48 nm, measured before (black solid lines) and after (red solid lines) exposure to 0.5 % H$_2$. A S/D bias of $V_{ds}$ = 50 mV was used for all measurements. Note that in (a), current plot for before H$_2$ exposure (i.e., measurement in dry air) was multiplied by 10 for clarity.

**Figure 3. QM thickness effects on sensor properties.** (a) Schematic representation of the sensing mechanism, showing the modulation of surface potential and surface electron scattering rates induced by the work function change of Pd NPs upon H$_2$ exposure. (b) Sensor response to 0.5% H$_2$ as a function of InAs QM thickness at $V_{gs}$=0V and $V_{ds}$=50 mV. Two orders of magnitude higher sensor response is observed in the 8 nm-thick InAs device as compared to the 48 nm-thick device. (c) Mobility change (left axis) and threshold voltage shift (right axis) upon exposure to H$_2$ for different InAs thicknesses.

**Figure 4. Quantum simulation of InAs QM sensors.** (a) and (b) The calculated band diagrams of InAs QMs with 8 and 48 nm thickness, respectively for both before and after H$_2$ exposure. The population percentage for each sub-band is also labeled. (c) and (d) The calculated carrier density profiles (left axis) and perturbation potential (right axis) for QMs with 8 and 48 nm thickness, respectively. For the carrier density profiles, both before (black lines) and after (red lines) $H_2$ exposure are shown. (e) and (f) Simulated current (a.u.) *versus* gate voltage characteristics for 8 nm and 48 nm-thick devices, respectively. The red (black) lines represent the device characteristics after (before) $H_2$ exposure. The simulation results are consistent with the experimental characteristics of Figure 2.

**References**

1. Law, M.; Kind, H.; Messer, B.; Kim, F.; Yang, P. D. *Angew. Chem. Int. Ed.* **2002**, *41*, 2405–2408.

2. Zhang, D.; Liu, Z.; Li. C.; Tang, T.; Liu. X.; Han, S.; Lei, B.; Zhou, C. *Nano Lett.* **2004**, *4*, 1919–1924.

3. Katz, E.; Willner, I. *ChemPhysChem* **2004**, 5,1084–1104.

4. Zheng, G.; Patolsky, F.; Cui, Y.; Wang, W. U.; Lieber, C. M.; *Nat. Biotech.* **2005**, *23,*1294-1301.

5. Peng, K. Q.; Wang, X.; Lee, S. T. *Appl. Phys. Lett.* **2009**, *95*, 243112.

6. Strelcov, E.; Lilach, Y.; Kolmakov, A. *Nano Lett*. **2009**, *9*, 2322–2326.

7. Seo, M.-H.; Yuasa, M.; Kida, T.; Huh, J.-S.; Shimanoe, K.; Yamazoe, N. *Sensors and Actuators B*. **2009**, 137, 513-520.

8. Wei, T. Y.; Yeh, P. H.; Lu, S. Y.; Wang, Z. L. *J. Am. Chem. Soc.* **2009**,*131*, 17690–17695.

9. Xu, C.; Tamaki, J.; Miura, N.; Yamazoe, N. *Sensors and Actuators B.* **1991**, 3, 147-155.

10. Yamazoe, N.; Miura, N. Some basic aspects of semiconductor gas sensor, *Chemical Sensor Technology* **1992**, 4, Kodansha and Elsevier, Tokyo.

11. Klemic, J. F.; Stern, E.; Reed, M.A. *Nat. Biotechnol.* **2001**, 19, 924–925.

12 Cui, Y.; Wei, Q.; Park, H.; Lieber, C. M. *Science* **2001**, 293, 1289–1292.

13. Fan, Z.; Wang, D.; Chang, P.; Lu, J. G. *Appl. Phys. Lett.* **2004**, 85, 5923.

14. Liao, L.; Lu, H. B.; Li, J. C.; He, H.; Wang, D. F.; Fu, D. J.; Liu, C.; Zhang, W. F. *J. Phys. Chem. C* **2007**, 111, 1900-1903.

15. Stern, E.; Wagner, R.; Sigworth, F. J.; Breaker, R.; Fahmy, T. M.; Reed, M. A. *Nano Lett.* **2007**, *7*, 3405-3409.

16. Yamazoe, N.; Shimanoe, K. *Sensors and Actuators B.* **2010**, 150, 132-140.

17. Takei, K. et al. *Nano Lett.* **2011**, 11, 5008-5012.

18. Lewis, F. A. The Palladium Hydrogen System, *Academic Press*, INC., **1967**, New York.

19. Kong, J.; Chapline, M.; Dai, H. *Adv. Mater.* **2001**, 13, 1384-1386.

20. Skucha, K.; Fan, Z.; Jeon, K.; Javey, A; Boser, B. *Sensors & Actuators*: B **2010**, 145, 232-238.

21. Ko, H. et al. *Nature* **2010**, 468, 286–289.

22. Datta, S. Quantum Transport: Atom to Transistor, *Cambridge University Press* **2005**, Cambridge.

23. Gámiz, F.; Roldán, J. B.; Cartujo-Cassinello, P.; López-Villanueva, J. A.; Cartujo, P. *J. Appl. Phys.* **2001**, 89, 1764-1770.

**Figure 1**

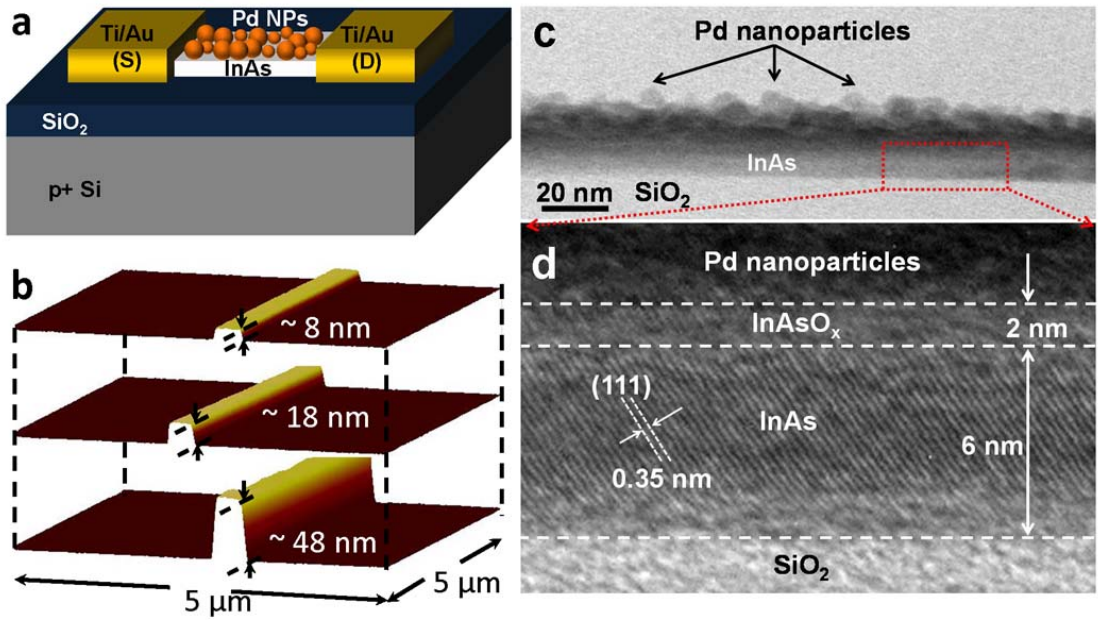

**Figure 2**

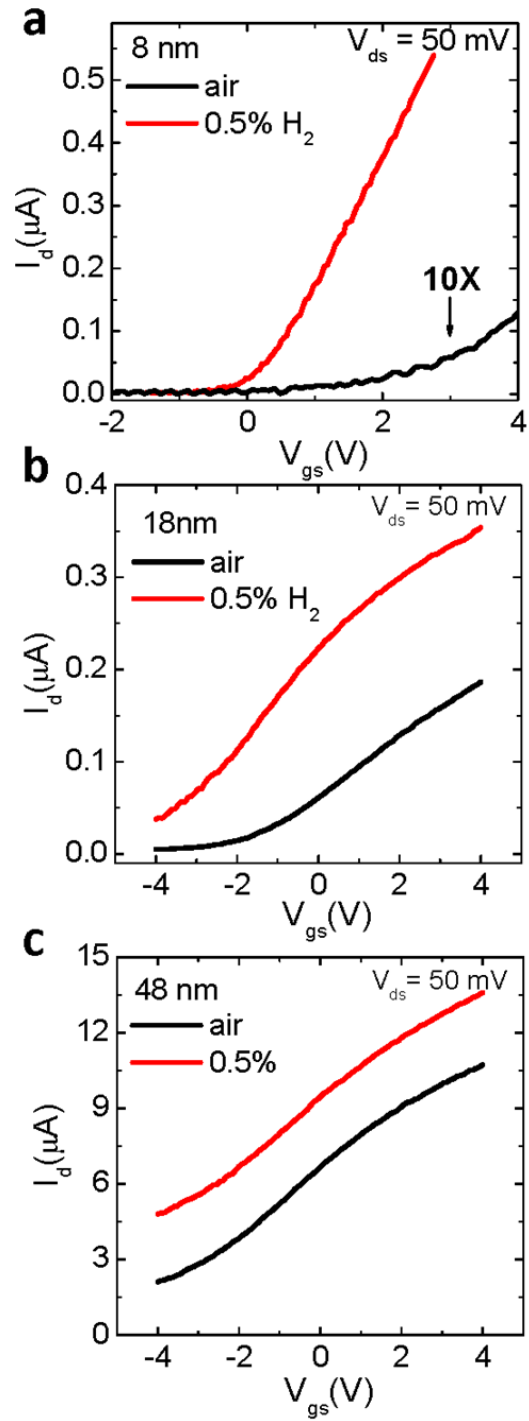

**Figure 3**

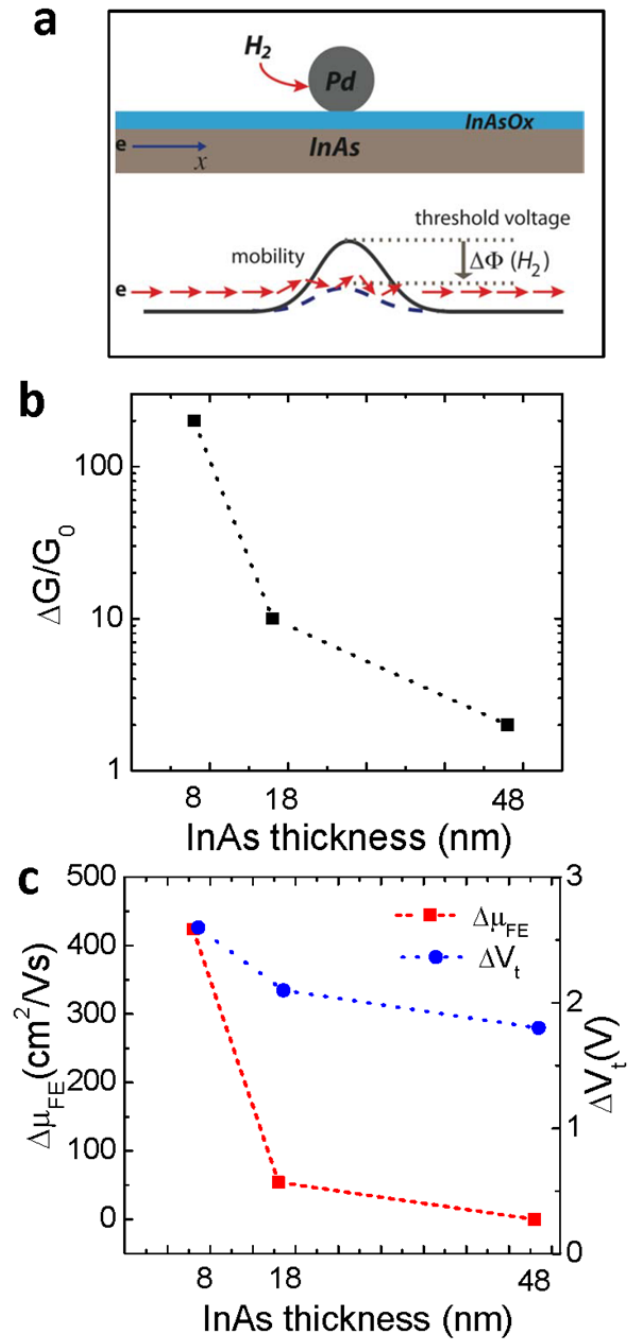

**Figure 4**

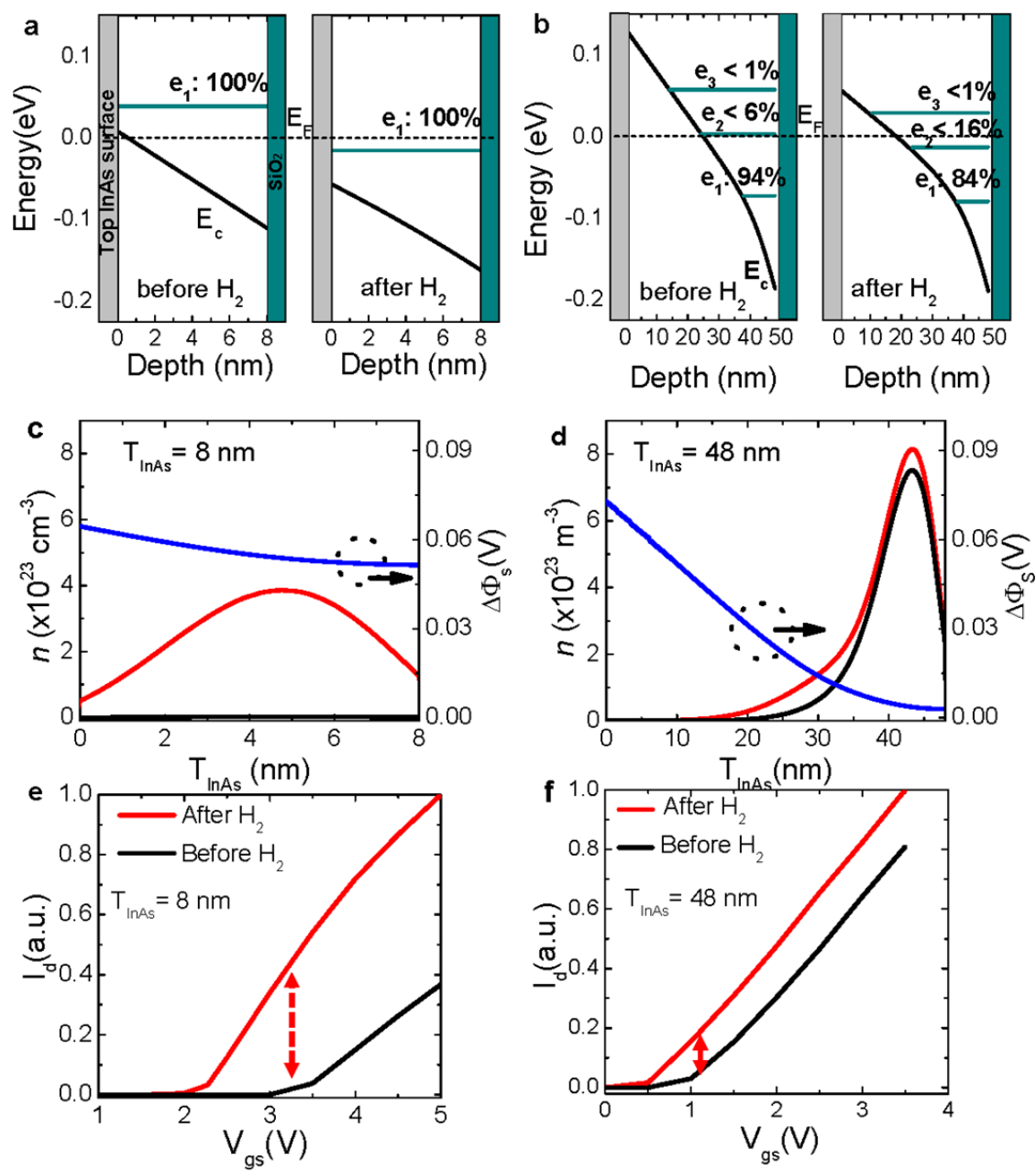

# TOC Figure

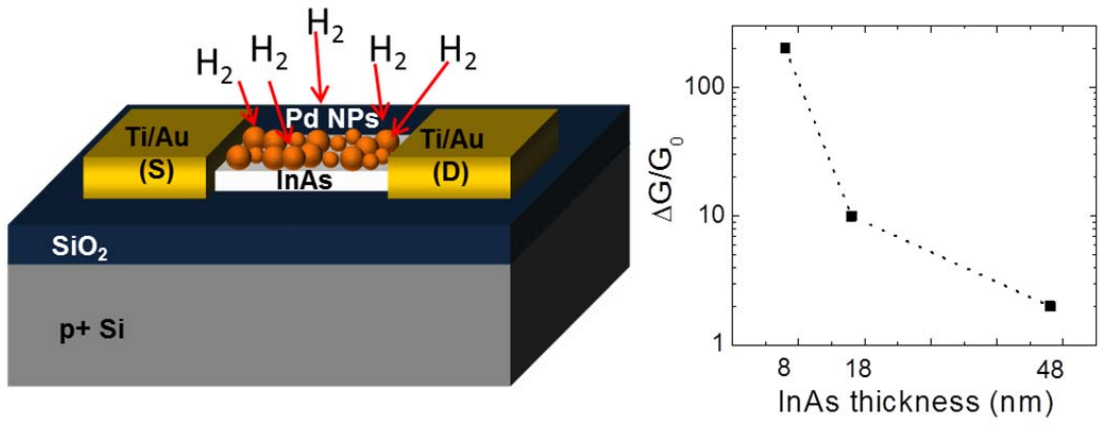

# SUPPLEMENTARY INFORMATION

# Quantum Size Effects on the Chemical Sensing Performance of Two-Dimensional Semiconductors

Junghyo Nah[1,2,3,†], S. Bala Kumar[4], Hui Fang[1,2,3], Yu-Ze Chen[5], Elena Plis[6], Yu-Lun Chueh[5], Sanjay Krishna[6], Jing Guo[4] and Ali Javey[1,2,3,*]

[1] Electrical Engineering and Computer Sciences, University of California, Berkeley, CA, 94720

[2] Materials Sciences Division, Lawrence Berkeley National Laboratory, Berkeley, CA 94720

[3] Berkeley Sensor and Actuator Center, University of California, Berkeley, CA, 94720

[4] Electrical and Computer Engineering, University of Florida, Gainesville, FL, 32611

[5] Materials Science and Engineering, National Tsing Hua University, Hsinchu 30013, Taiwan, ROC, 30013

[6] Electrical and Computer Engineering, University of New Mexico, Albuquerque, NM, 87106

* Corresponding author: ajavey@eecs.berkeley.edu

**Electrical characteristics change after palladium functionalization**

Figure S1 shows the transfer ($I_d$-$V_{gs}$) characteristics of back-gated InAs XOI devices with different InAs thicknesses for both as-fabricated (black solid line) and after Pd-coating (red solid line) for the same devices. Figure S1 (a), (b), and (c) correspond to the device characteristics of 8 nm, 18 nm, and 48 nm InAs XOI, respectively. The $I_d$-$V_{gs}$ characteristics were measured at $V_{ds}$ = 50 mV. Before Pd-functionalization of InAs surfaces, the highest ON/OFF current ratio is observed in the 8 nm-thick InAs XOI, indicating better electrostatic control by the back-gate biases by comparison to thicker InAs XOI devices. After Pd-functionalization of the InAs surfaces, there was loss of gate-control in all devices due to screening effects by deposited Pd layer resulting in current reduction. The most obvious characteristics change was observed in the 8 nm-thick InAs XOI device, where the currents were reduced by two orders of magnitude after Pd deposition. These results clearly indicate that the potential inside the 8 nm-thick InAs XOI device is more strongly affected by Pd-functionalization of the surface by comparison to other devices.

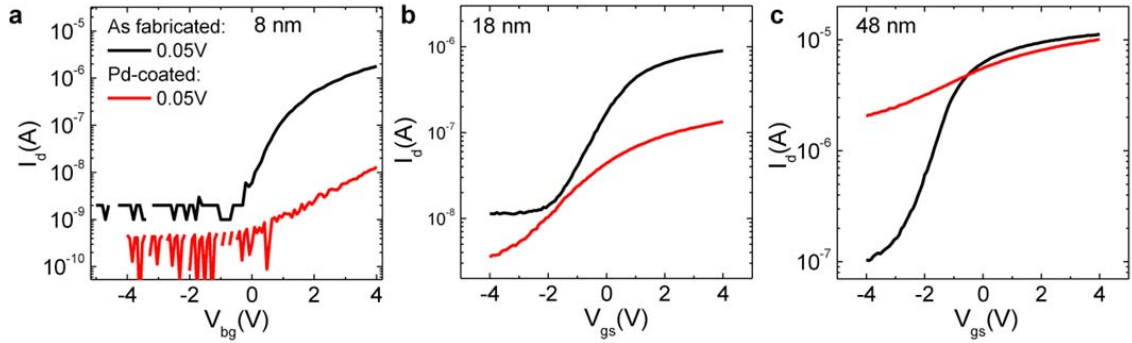

**Figure S1. Palladium functionalization impacts on transfer characteristics ($I_d$-$V_{gs}$) of back-gated InAs XOI device** (a) Transfer characteristics of 8 nm-thick InAs XOI (b) 18 nm-thick InAs XOI (c) 48 nm-thick InAs XOI, measured at $V_{ds}$ = 50 mV. The black solid lines and red solid lines represent the device characteristics before/after palladium coating, respectively.

**Aggregation of evaporated Pd film into nanoparticles**

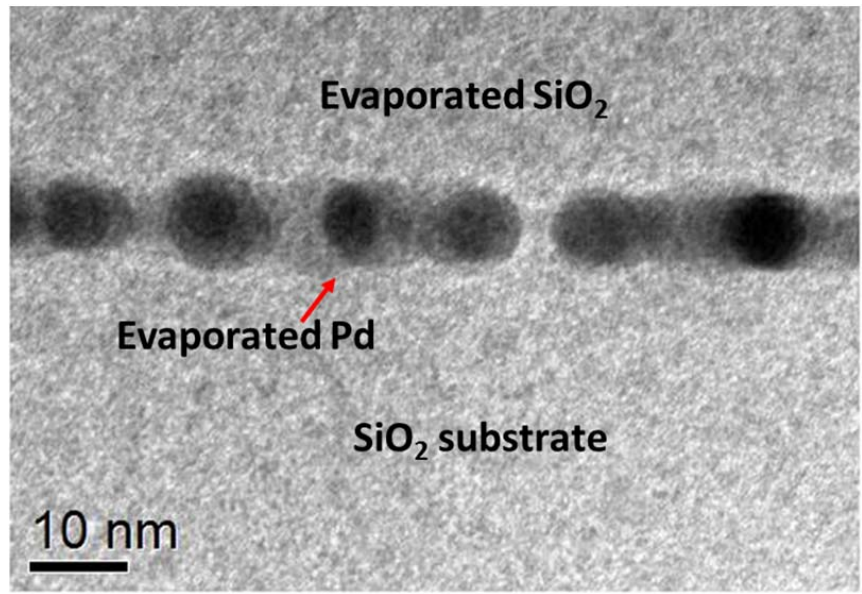

**Figure S2. TEM image of evaporated Pd layer (thickness, ~ 2nm) on a SiO$_2$ substrate.** It can be clearly seen that the evaporated Pd film is aggregated into Pd nanoparticles (~10nm)

## Stability of gas sensor after multiple cycle

Next, to ensure the stabilities of the sensor devices, the sensors were exposed to 0.5% $H_2$ for 12 s at each cycle, which were repeated in multiple cycles after its recovery to the initial state [Figure S3 (a), (b), (c)]. Here, the sensor response is defined as $\Delta G/G_0$, where $G_0$ is the initial conductance in ambient air and $\Delta G$ is the change of the conductance from $G_0$ after $H_2$ exposure. For all devices, repeating sensor responses were observed at each 0.5% $H_2$ exposure cycle. In addition, sensor responses to $H_2$ exposures were achieved within 1 s in all the devices. As expected, the sensor response in a 8 nm-thick InAs XOI gas sensor was at least one order of magnitude higher than one observed in the 48 nm-thick InAs XOI gas sensor.

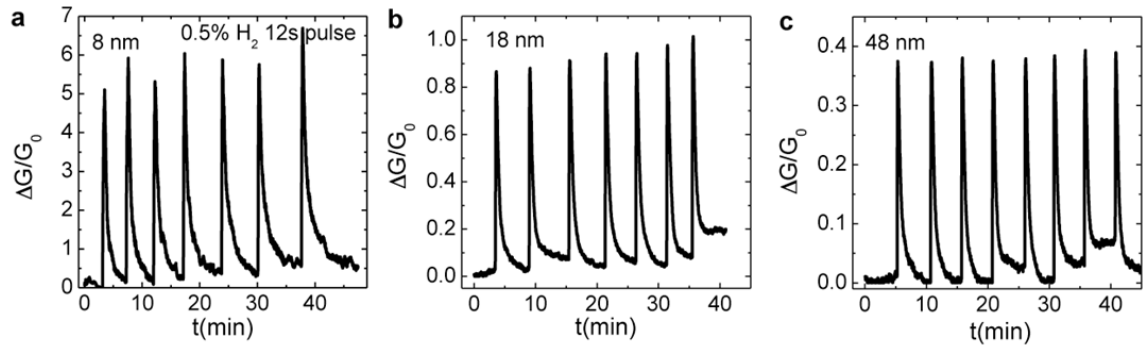

**Figure S3. Sensor responses of palladium-functionalized InAs gas sensor to 0.5 % hydrogen exposure**. Sensor response ($\Delta G/G_0$) of (a) 8 nm- (b) 18 nm- (c) 48 nm-thick InAs XOI gas sensor. Devices were repeatedly exposed to 0.5 % $H_2$ for multiple cycles of 12s at constant $V_{ds} = 50$ mV. Note that here, the exposure pulse was short and the response did not reach the saturation value. This is in contrast to the Figure 2 studies, where the exposure to $H_2$ was performed for 10 min to reach the equilibrium state. Therefore, the response magnitude here is smaller than that of Fig. 2.

**Sensor response at different $H_2$ concentrations (8 nm- and 18 nm-thick InAs)**

Sensor response of the 8 nm and 18 nm thick-XOI gas sensor were measured at different concentrations of $H_2$ gas (diluted in dry air) as shown in Figure S4. The results show that sensor response monotonically increases as the $H_2$ concentration increases for the explored concentration range. Consistent with the Figure S3 and Figure 2, much higher sensor response was obtained in the 8 nm-thick InAs XOI gas sensor.

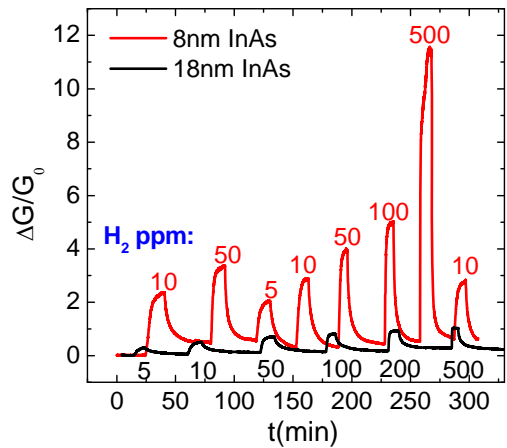

**Figure S4.** Sensor response of Pd-functionalized InAs XOI gas sensor with different thicknesses, measured at various $H_2$ concentrations.

**Electrostatic simulations for InAs XOI gas sensor**

Here we use a self-consistent Poisson-Schrödinger solver[1] to obtain the potential profile across the depth of the channel, the eigenenergies, wavefunction and the charge density. In the self-consistent Poisson-Schrodinger solver, electron interaction is treated under Hartree approximation. The sensor operates at room temperature, and the carrier density is in the typical range as that of conventional semiconductor on insulator transistors. Many body effects, such as exchange and correlation, are not expected to play an important role. For the simulation, InAs XOI gas sensors were modeled as double-gated MOS capacitors, composed of top-gate electrode (Pd), native gate oxide (2 nm, $\varepsilon = 2.2$), InAs film (8, 18, 48nm), back-gate oxide (50 nm, $\varepsilon = 3.9$), and back-gate electrode (Si ($n+$)), which is similar to the actual sensor structures [Fig. S5b(inset)]. In the calculation, top-gate voltage ($V_{tg}$) is floated to simulate the device before Pd deposition, $V_{tg} = -0.2$V is applied to represent the device after Pd layer deposition, $V_{tg} = -0.1$V corresponds to device operation under $H_2$ exposure, and $V_{gs} = -3.0$V is applied. Figure S5 shows the electrostatic simulations of current density for different thicknesses of InAs. The results show that threshold voltage ($V_t$) shift is increased as the channel thickness scales down.

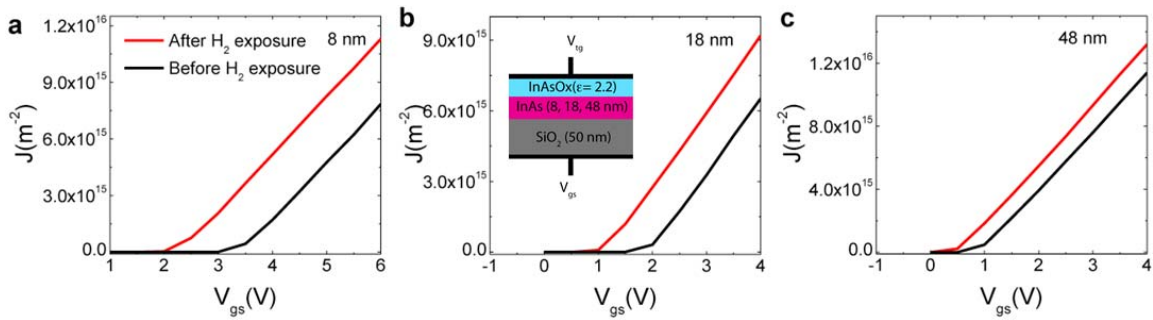

**Figure S5. Electrostatic analysis of carrier density change in $H_2$ for different InAs thicknesses.** (a) 8 nm-thick InAs XOI (b) 18 nm-thick InAs XOI (c) 48 nm-thick InAs XOI.

**Calculation of potential disorder scattering (PDS) mobility**

The random deposition of Pd, creates an uneven potential profile along the InAs channel, giving rise to the PDS. Here, we use a similar concept of modeling the surface roughness scattering[2] to model the PDS; since the surface roughness scattering is also model based on the uneven potential profile along the channel. Thus we defined PDS mobility as $\mu_{PDS} = (U^2\mu_{PDS0})\{\Sigma |(\Psi_v|\Phi_i|\Psi_\mu)|^2 P_v P_\mu\}^{-1}$, where $\Psi_n$ is the wave-function occupying the $n^{th}$ state, and $P_n$ is the percentage of the electrons occupying the $n^{th}$ state, $\Phi_i$ is perturbation potential and $U^2\mu_{PDS0}$ is a fitting parameter. The perturbation potential, $\Phi_i=|V_i-V_0|$, where $V_i(V_0)$ is the potential across the depth of the InAs channel when a top gate voltage of $V_{tg}$ (floating gate) is applied. Thus, the perturbation potential change after $H_2$ exposure is defined as $\Delta\Phi_s =| \Phi_2 - \Phi_1 |$, where $\Phi_2$ and $\Phi_1$ are the perturbation potential before/after $H_2$ exposure, respectively. Based on the obtained $\mu$ values, the source-drain current characteristics were qualitatively estimated as $I_d \sim N\mu$, where $N$ is the total charge density and $\mu$ is the mobility of the device. Here, $\mu$ is determined using Matthiessen's rule, $\mu^{-1} \sim \mu^{-1}_{PDS} + \mu_0^{-1}$, where $\mu_{PDS}$ is the PDS mobility and $\mu$ is the total mobility due to other mechanism. In thin InAs films, the effect of PDS is most predominant and $\mu$ can be approximated as $\mu_{PDS}$.

**Semiclassical simulation results of the 8 nm thick InAs**

Figure S6 shows the semiclassical simulation results of carrier distribution and perturbation potential change ($\Delta\Phi_s$) upon $H_2$ exposure. It demonstrates small $\Delta\Phi_s$ in comparison to quantum simulation results in Fig. 4(c).

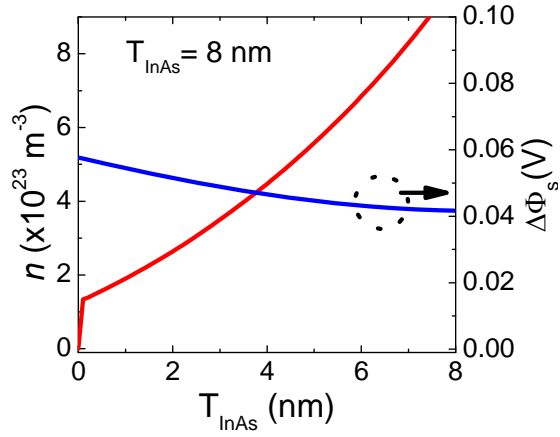

**Figure S6. Semiclassical calculation of the carrier concentration and perturbation change in the 8 nm-thick InAs QM upon $H_2$ exposure.**

## Simulation results for a 8 nm-thick Si

We also investigated the 8 nm-thick Si membranes using the same approach [Fig. S7]. Compared with InAs (8 nm), the charge centroid of Si (8nm) is located closer to the $SiO_2$ interface, and the value of $\Delta\Phi_s$ is relatively small and decreases more rapidly from the top Si surface. This can be attributed to the larger effective mass ($m^*$~$0.98m_0$) and smaller Bohr's radius (~ 4.5 nm) of Si compared with InAs. As a result, the current *vs*. voltage characteristics show only the $V_t$ shift after $H_2$ exposure without slope (transconductance) changes, different from the same thickness of InAs. These results clearly indicate the advantage of using InAs for the study of size effects down to quantum confinement limits.

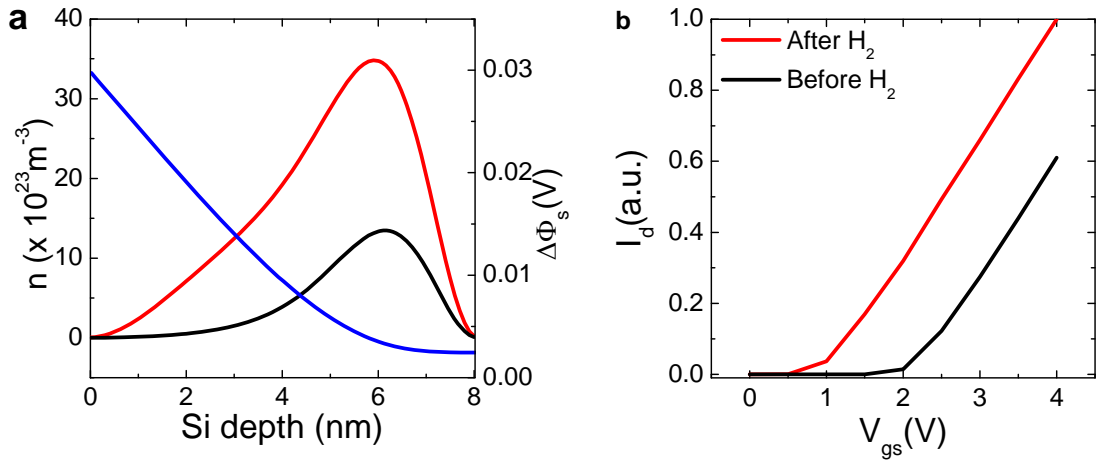

**Figure S7. Carrier density change and potential disorder in $H_2$ for a 8 nm thick Si device.** (a) carrier density change (right axis) and potential disorder (blue line, right axis) upon $H_2$ exposure (b) current vs. voltage characteristics of 8 nm-thick Si device.

## References

[1] Datta,S. Quantum Transport: Atom to Transistor, *Cambridge University Press*, Cambridge, **2005**.

[2] Gámiz, F.; Roldán, J. B.; Cartujo-Cassinello, P.; López-Villanueva, J. A.; Cartujo, P. *J. Appl. Phys.* **2001**, 89, 1764-1770.